\documentclass[twocolumn]{webofc}

\usepackage[varg]{txfonts}   % Web of Conferences font
\usepackage{hyperref}
\usepackage{url}
\usepackage{physics}
\hypersetup{colorlinks=true,citecolor=blue,urlcolor=blue,linkcolor=blue}
\begin{document}

\title{
Correcting for neutron width fluctuations in Hauser-Feshbach gamma branching ratios
}

% POSTER PRESENTATION: 3 PAGES LENGTH LIMIT
\author{\firstname{Oliver} \lastname{Gorton}\inst{1}\fnsep\thanks{\email{gorton3@llnl.gov}} \and
        \firstname{Jutta} \lastname{Escher}\inst{1}\fnsep
}

\institute{Lawrence Livermore National Laboratory}

\abstract{
Porter-Thomas fluctuations of neutron widths skew compound nuclear decay
probabilities from their statistical Hauser-Feshbach values. We present a
straightforward method to correct Hauser-Feshbach calculations for these
fluctuations, useful for modeling near-threshold competition between gamma and
neutron emission following beta decay or when standard width fluctuation
corrections are inadequate.} 

\maketitle

Beta decay plays a central role in nucleosynthesis and understanding beta decay
rates is one of the central tasks of the nuclear theory community and the new
Facility for Rare Isotope Beams
(FRIB)~\cite{arcones2017white,schatz2022horizons}; it is also a mechanism to
study short-lived isotopes~\cite{spyrou2014novel}. In neutron-rich systems, beta
decay products may undergo beta-delayed neutron emission (BDNE), in which case
the gamma-channel branching ratio (as a function of the product excitation
energy $E_x$) becomes an important quantity:  
\begin{equation}\label{eq:qoi}
    \left \langle \frac{\Gamma_{\gamma: i}}
    {\Gamma_{\gamma:i}+\Gamma_{n:i}}\right \rangle
    = 
    \left \langle \frac{\sum_{f=1}^{k_\gamma}\Gamma_{\gamma: fi}}
    {\sum_{f=1}^{k_\gamma}\Gamma_{\gamma: f'i}
    + \sum_{f=1}^{k_n}\Gamma_{n: fi}}\right \rangle,
\end{equation}
where $\Gamma_{\gamma: fi}$ is the partial decay width for a gamma transition
from a level $i\to f$, and similarly for the partial neutron decay widths
$\Gamma_{n: fi}$. ($\Gamma = \hbar T$ for a transition probability $T$.) The
sums extend over all $k$ final states allowed by energy, angular momentum, and
parity rules. The average extends over all initial states in some initial energy
bin. In principle, equation~\eqref{eq:qoi} can be calculated with a
Hauser-Feshbach (HF) reaction code. However, HF theory assumes that:
\begin{equation}\label{eq:hfapprox}
    \left \langle \frac{\Gamma_{\gamma: i}}{\Gamma_{\gamma:i}+\Gamma_{n:i}}\right \rangle
    \approx \frac{\langle\Gamma_{\gamma: i} \rangle}{\langle \Gamma_{\gamma:i}
    \rangle+\langle\Gamma_{n:i}\rangle}.
\end{equation}
In contradiction, the authors of Ref.~\cite{valencia2017total} found that
HF theory greatly under-predicts the observed gamma branching
ratio. Valencia et al.~\cite{valencia2017total} found that a significant
improvement over the HF result is obtained by including the effects of
Porter-Thomas fluctuations of the neutron partial widths. They implemented this
effect with a custom program similar to {\tt
DICEBOX}~\cite{becvar1998simulation}, but including the effects of neutron width
fluctuations in addition to gammas.

In these proceedings, we study how neutron width fluctuations affect the gamma
emission branching ratio. We confirm that the gamma channel branching ratio can
be significantly enhanced relative to the statistical prediction when there are
few final states in the neutron exit channel. Such fluctuation effects are not
included in standard HF codes. We therefore propose a correction
factor that is inspired by the width fluctuation correction (WFC)
factor~\cite{moldauer1980statistics, moldauer1976evaluation,
hilaire2003comparisons} which allows one to avoid the costly Monte Carlo cascade
calculations demonstrated in Ref.~\cite{valencia2017total}.

Porter-Thomas theory states that the partial decay widths $\Gamma_{fi} \propto |
\langle \Psi_f | \hat{\mathcal{O}} | \Psi_i \rangle |^2$ between initial states $| \Psi_i \rangle$ and final states $| \Psi_f \rangle$ can be considered as
random numbers following a chi-squared distribution with one degree of freedom, and that therefore the total decay width for an initial level $i$ with $k$ partial widths, $\Gamma_i = \sum_{f=1}^k \Gamma_{fi}$, follows a chi-squared distribution with $k$ degrees of
freedom~\cite{moldauer1976evaluation}. The corresponding probability density
function for $\Gamma_{fi}$ can be written: $P(x, k) = G(r)^{-1}r(rx)^{r-1}e^{-rx},$ where $x
= \Gamma_{fi}/\langle \Gamma_{fi}\rangle$ is the partial width normalized to its
mean, $G(r)$ is the gamma-function, and $r=k/2$. 
It is not obvious that Porter-Thomas fluctuations of neutron widths will
increase the average gamma branching ratio relative to the HF prediction given
by Eq.~\eqref{eq:hfapprox}. To illustrate the effect, we conducted a numerical
experiment by calculating the gamma branching Eq.~\eqref{eq:qoi} for randomly
generated partial widths. In the first round of simulations, we arbitrarily
assume that $\langle \Gamma_n \rangle = \langle \Gamma_\gamma \rangle$. We relax
this assumption later. We work in units of the averages so that $P(x, k)$ given above
applies directly to the partial widths ($x = \Gamma_{fi}$). Centrally important
is the assumption that the neutron total widths include only a few terms $k$, so
that $P(\Gamma_n) = P(\Gamma_n, k)$. We further assume the gamma total widths
include many terms (owing to the high excitation energy required for neutron
emission), so that $P(\Gamma_\gamma) \approx P(\Gamma_\gamma, k=\infty) \approx
\delta(\Gamma_\gamma-1)$. With these assumptions, the purely HF
estimate of $\langle \Gamma_\gamma / \Gamma_\text{total} \rangle$ for any $k$
neutron partial widths is always:
\begin{equation}
    \frac{\langle\Gamma_\gamma\rangle}
    {\langle\Gamma_\gamma\rangle+\langle\Gamma_n\rangle} 
    = \frac{1}{1+\langle P(\Gamma_n, k) \rangle } = \frac{1}{2},
\end{equation} 
regardless of the number of neutron partial widths. Next, we numerically
simulate the ``true'' gamma branching ratio by generating random values of the
neutron total widths $\Gamma_{n:i}$ from the appropriate chi-squared
distribution $P(x, k)$ and computing the exact branching ratio:
\begin{equation}
    \frac{\Gamma_\gamma}{\Gamma_\text{total}} 
    = \frac{1}{1 + \Gamma_n},
\end{equation}
After generating $n=10^6$ samples of the neutron widths and exact branching
ratios, we compute the mean ratio $\langle
\Gamma_\gamma/\Gamma_\text{total}\rangle$ of all the samples. The final results
are relatively insensitive to the number of samples, but we use a large number
to produce smooth histograms. 

\begin{figure}[hb]
    \centering
    \includegraphics[width=.8\columnwidth, trim={0.1cm 0.2cm 0.1cm 0.0cm}, clip]{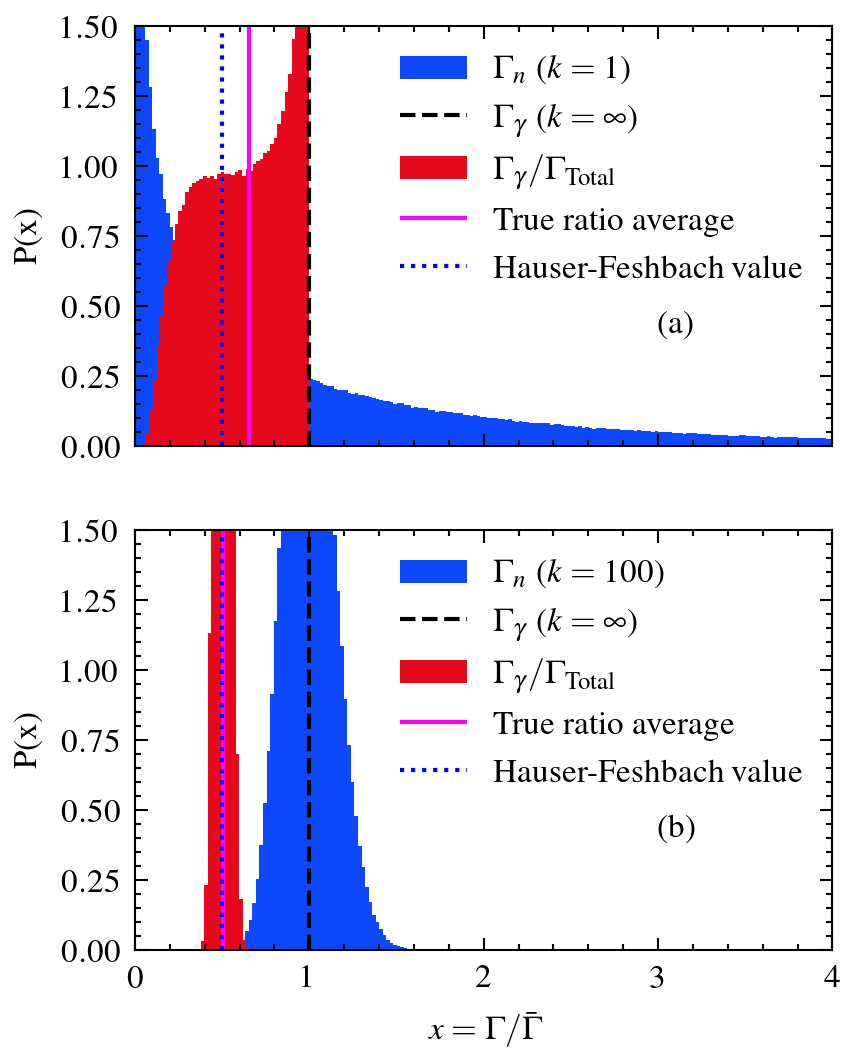}
    \caption{Porter-Thomas fluctuation toy model wherein the average neutron and
    gamma widths are equal. See text for discussion.}
    \label{fig:ptsim1}
\end{figure}

The results of the numerical simulations for $k=1$ and $k=100$ are shown in
Figure~\ref{fig:ptsim1} panels (a) and (b), respectively. The gamma total widths
are constant (black dashed line) while the neutron total widths are randomly
distributed (blue histograms). The resulting distribution of width ratios are
shown in the narrow red histograms.
At $k=1$ we observe the maximum effect of PT fluctuations. We obtain $\langle
\Gamma_\gamma/\Gamma_\text{total}\rangle=0.66$. As anticipated, the gamma branch
is enhanced with respect to the HF prediction of 0.5; the increase is about 33
percent. From the $k=100$ simulation, $\langle
\Gamma_\gamma/\Gamma_\text{total}\rangle=0.503$, which is close to the HF
prediction. As expected, the effect of PT fluctuations are suppressed as the
number of neutron partial widths increases. 

We have shown how Porter-Thomas fluctuations of the neutron partial widths can
enhance the gamma branching ratio. In the second round of simulations, we
explore how the enhancement depends on the number of partial widths and the size
of the HF estimate. We varied the number of neutron partial widths from $k=1$ to
$k=100$ and relaxed the arbitrary assumption that $\langle \Gamma_n \rangle =
\langle \Gamma_\gamma \rangle$. To preserve the generality of our findings, we
normalize the average gamma width relative to the average neutron width. We set
$P(\Gamma_\gamma) \approx \delta(\Gamma_\gamma-g)$ for some constant $g$ in
units of the average neutron total widths $\langle \Gamma_n \rangle$. The
simulated branching ratios become:
\begin{equation}\label{eq:exsim}
    \frac{\Gamma_\gamma}{\Gamma_\text{total}} 
    = \frac{g}{g + \Gamma_n},
\end{equation}
where again the neutron total widths are randomly drawn from $P(\Gamma_n, k)$.
Since $\langle \Gamma_n\rangle = 1$ by construction, changing $g$ is equivalent
to changing the ratio
\begin{equation}\label{eq:relwidth}
    y \equiv \frac{\langle \Gamma_\gamma \rangle}
    {\langle\Gamma_\gamma\rangle+\langle \Gamma_n\rangle}
    = \frac{g}{g+1},
\end{equation}
which is the HF estimate of the gamma branching ratio. We simulate HF gamma
branching ratios between $y=10^0$ and $y=10^{-5}$ to span the range
encountered in Ref.~\cite{valencia2017total}. 
To model the impact of the neutron width fluctuations, we define a
Porter-Thomas width fluctuation correction (PT WFC) factor, which relates the
exact gamma branching ratio computed with Eq.~\eqref{eq:exsim} to the HF
estimate, Eq.~\eqref{eq:relwidth}:
\begin{equation}\label{eq:PTWFCF}
    W(k, y) 
    \equiv \frac{\langle \Gamma_\gamma/\Gamma_\text{total}\rangle}
    {\langle\Gamma_\gamma\rangle
    /(\langle\Gamma_\gamma\rangle+\langle\Gamma_n\rangle)}
    = \frac{\text{True ratio}}{\text{HF estimate y}}.
\end{equation}
Importantly, this correction factor is independent of the absolute value of
either the average neutron decay width or average gamma decay width; it depends
only on the number of neutron partial widths $k$ and the HF gamma
branching ratio $y$.
% This correction factor is completely analogous to the correction factor of
%Moldauer~\cite{moldauer1976evaluation} and we expect it can be applied to
%correct HF codes for this effect.

\begin{figure}[ht]
    \centering
    \includegraphics[width=.8\columnwidth, trim={0.1cm 0.1cm 0.1cm 0.0cm}, clip]{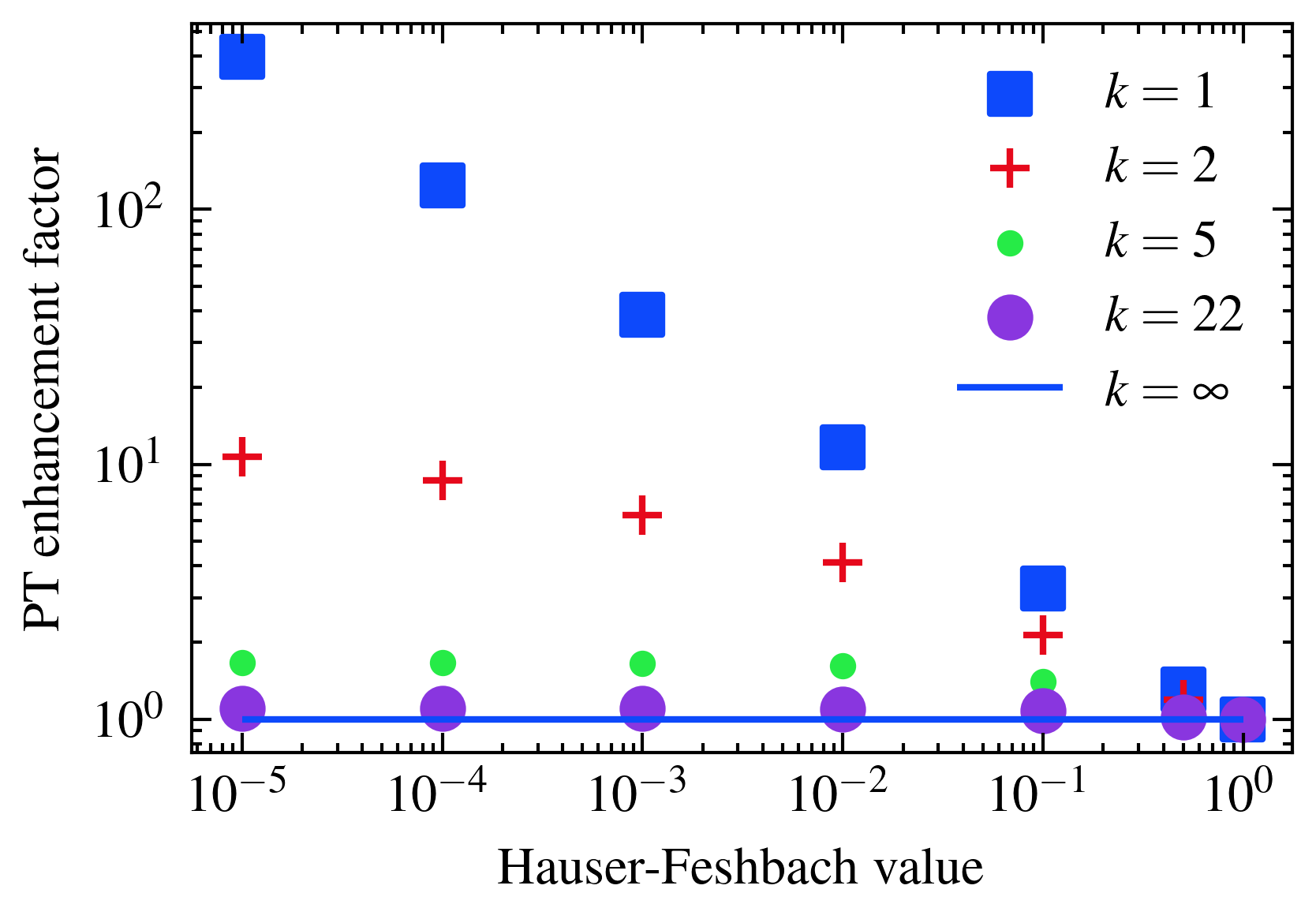}
    \caption{Porter-Thomas (PT) correction factor versus HF
    prediction $y$ for different numbers of neutron partial widths $k$. }
    \label{fig:pttherm}
\end{figure}

Figure~\ref{fig:pttherm} shows the smooth decay of the PT WFC factor,
Eq.~\eqref{eq:PTWFCF}, from its maximum at $k=1$ where the PT enhancement
can be up to two orders of magnitude at $y=10^{-5}$. By $k=5$, all curves are
below an enhancement of 2. 
For all curves to be below an enhancement of 1.1, one requires $k>22$. We find
that the enhancement factor follows a smooth and systematic trend. This effect
is independent of any energy dependence of the absolute strength of the neutron
partial widths (which are known to have $\sqrt{E}$
dependence~\cite{weidenmuller2010distribution}) and depends only on the number
of neutron partial widths $k$ and the gamma branching ratio $y$ given by
Eq.~\eqref{eq:relwidth}. We can therefore compute the correction factor $W(k,
y)$ \textit{a priori} and apply it to our HF estimate to
approximate the true ratio $\langle \Gamma_\gamma/\Gamma_n \rangle$.

\begin{figure}[h]
  \centering
  \includegraphics[width=\columnwidth, trim={0.1cm 0.15cm 0.1cm 0.0cm}, clip]{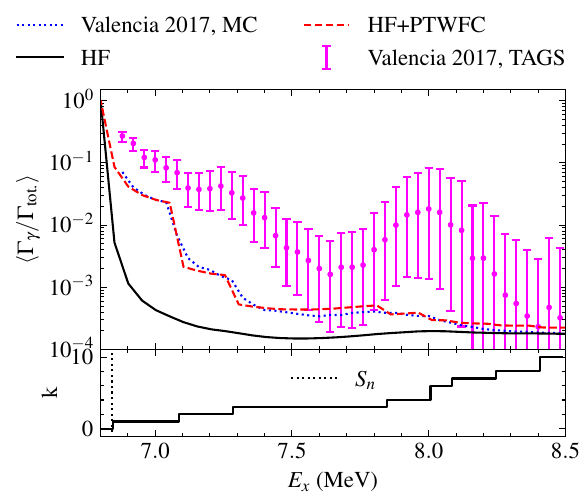}
  \caption{Gamma branching ratio with and without the PTWFC
  factor~\eqref{eq:PTWFCF}. The lower panel shows the number of neutron partial
  widths $k$ available at each excitation $E_x$. The discontinuities in the
  HF+PTWFC calculation line up with changes in
  $k$.}\label{fig:rb94bdne-corrected}
\end{figure}

Figure~\ref{fig:rb94bdne-corrected} shows an application of the PT WFC to
beta-delayed gamma emission from Ref.~\cite{valencia2017total}. We show our
original HF calculation (HF) which used gamma strength functions from
Ref.~\cite{goriely2019reference, goriely2018gognyhfb}, the corrected calculation
using the $W(k,y)$ correction factor (HF+PTWFC), and the Monte-Carlo cascade
simulation (Valencia 2017, MC) from Ref.~\cite{valencia2017total}. All three
consider only those decays from $J^\pi=3^-$ states. At each excitation energy we
determine $y$ from the HF calculation, then apply the correction
Eq.~\eqref{eq:PTWFCF} from Figure~\ref{fig:pttherm}. $k$ is equal to the
cumulative number of levels in the residual nucleus available for neutron
emission. 
%$y$ is equal to the original HF ratio of average widths. 
We reproduce the same enhancement produced by the Monte Carlo decay simulation,
within some margin of error attributable to differences in the details of the
nuclear level densities and gamma strength functions used.

In conclusion, we have shown that a simple Moldauer-type correction factor can
be applied to correct standard HF calculations for the effects of neutron width
fluctuations. The correction is broadly applicable to branching ratios near
particle emission threshold and is easy to implement, enabling adaptation of
existing HF codes without expensive Monte Carlo simulations.

\textit{Prepared by LLNL under Contract DE-AC52-07NA27344, with support from a
WPD ACT award. We thank J. Tain for sharing with us the data from
Ref.~\cite{valencia2017total}.}

\bibliography{library} 

\end{document}